\documentstyle[11pt]{article}

\def\spose#1{\hbox to 0pt{#1\hss}}
\def\lta{\mathrel{\spose{\lower 3pt\hbox{$\mathchar"218$}}
     \raise 2.0pt\hbox{$\mathchar"13C$}}}
\def\gta{\mathrel{\spose{\lower 3pt\hbox{$\mathchar"218$}}
    \raise 2.0pt\hbox{$\mathchar"13E$}}}

      \def\Ov{{\mit\Omega}_{\rm v}}

 \def\Gam{{\mit\Gamma}}     \def\GeV{\,{\rm GeV}}

 \def\xis{\xi_{\sigma}}     \def\xia{\xi_\star}       \def\xif{\xi_{\rm f}}
           \def\zf{{\cal Z}_{\rm f}} \def\z{{\cal Z}}
 \def\ms{m_\sigma}          \def\mx{m_{\rm x}}        \def\mc{m_{\rm c}}
 \def\mP{m_{\rm P}\!}       \def\mp{m_{\rm p}}          
 \def\Ev{E_{\rm v}}         \def\lv{\ell_{\rm v}}     
 \def\nf{n_{\rm f}}         \def\nv{n_{\rm v}}        
 \def\nua{\nu_\star}        \def\nuf{\nu_{\rm f}}
 \def\rhv{\rho_{\rm v}}     \def\rhb{\rho_{\rm b}}    
 \def\rhc{\rho_{\rm c}}     \def\rhN{\rho_{_{\rm N}}}
 \def\aa{{g^*}}                     
 \def\aas{{g^*_\sigma}}     \def\aaf{{g^*_{\rm f}}}

 \def\q{\zeta}              \def\p{\varepsilon}       \def\ef{\varepsilon}
       
           \def\T{{\cal T}}          \def\Th{T}   
            
 \def\Tx{T_{\rm x}}       \def\Ts{T_\sigma}     \def\Tf{T_{\rm f}} 
 \def\Tr{T_{\rm r}}      
 \def\Ta{T_\star}         \def\Tra{T_\dagger}   
 \def\TN{T_{_{\rm N}}}    
 \def\TGUT{T_{_{\rm GUT}}}

\begin{document}

\newcommand{\vp}{\varphi}
\newcommand{\be}{\begin{equation}} \newcommand{\ee}{\end{equation}}
\newcommand{\bea}{\begin{eqnarray}} \newcommand{\eea}{\end{eqnarray}}

\footnotesep=8pt
\begin{center}
{\Large\bf CHIRAL VORTONS AND COSMOLOGICAL CONSTRAINTS ON PARTICLE PHYSICS
\footnote{PACs:{\tt 98.80.Cq, 11.27.+d}}}\\
\baselineskip=16pt
\vspace{0.75cm}

{\bf Brandon Carter\footnote{Electronic address: 
{\tt Brandon.Carter@obspm.fr}} and 
Anne-Christine Davis\footnote{Electronic 
address: {\tt a.c.davis@damtp.cam.ac.uk}} }\\
\vspace{0.4cm}
{\em ~\\$^1$ D.A.R.C., C.N.R.S, Observatoire de Paris-Meudon, 
92 195 Meudon, France.}{\em ~\\$^2$Department of Applied 
Mathematics and Theoretical Physics\/},\\ 
{\em University of
Cambridge\/},\\{\em Cambridge CB3~9EW, United Kingdom}\vspace*{0.75cm}\\
\end{center}
\baselineskip=20pt
\begin{center}
  {\bf\large Abstract}
\end{center}
\vspace{0.2cm} {\baselineskip=10pt} We investigate the cosmological
consequences of particle physics theories that admit stable loops of
current-carrying string - vortons. In particular, we consider chiral
theories where a single fermion zero mode is excited in the string
core, such as those arising in supersymmetric theories with a D-term.
The resulting vortons formed in such theories are expected to be more
stable than their non-chiral cousins. General symmetry breaking
schemes are considered in which strings formed at one symmetry
breaking scale become current-carrying at a subsequent phase
transition. The vorton abundance is estimated and constraints placed
on the underlying particle physics theories from cosmological
observations. Our constraints on the chiral theory are considerably
more stringent than the previous estimates for more general theories.
\vspace*{8pt}
\noindent

\renewcommand{\thefootnote}{\arabic{footnote}}
\setcounter{footnote}{0}
\newpage
\baselineskip=24pt

\section{Introduction}

There is considerable interest in supersymmetric theories, both from a
theoretical and phenomenological viewpoint. Indeed, current experiments
seem to suggest that the inclusion of supersymmetry is necessary for the
couplings to unify in a grand unified theory and to solve the hierarchy 
problem. Supersymmetry also underpins
superstring theory. It is thus important to explore the cosmological
consequences of such theories.

Recently it was shown that many supersymmetric theories admit cosmic
strings \cite{DDT1,DDT2}. The microphysics of such strings were investigated
and it was shown that they necessarily result in fermion zero modes in the
string core. The importance of this result is that the fermion zero modes
can be excited and move along the string. This completely changes its
properties, resulting in it becoming a current-carrying string
\cite{Witten85}. Both abelian \cite{DDT1} and non-abelian \cite{DDT2} theories 
were considered, and similar results found in both cases. For the abelian
case symmetry breaking by both an F-term and a D-term were considered. For
the theory with an F-term it was found that there is a pair of fermion 
zero modes, one left mover and one right mover. However, in the D-term 
case there was a single fermion zero mode, either a left mover or a right 
mover.

Since supersymmetry is not observed in nature the effect of supersymmetry
breaking was also considered \cite{DDT2}. In most cases supersymmetry
breaking enables the left and right movers to mix, destroying the zero modes, 
However, for the D-term theory the chiral zero mode survives supersymmetry 
breaking because there is no other zero mode for it to mix with. This theory 
is of importance because many superstring compactifications result in
a $U(1)$ symmetry being broken with a D-term, hence giving rise to the
cosmic strings considered in \cite{DDT1}. However, the presence of
fermion zero modes completely changes the cosmology of such strings and
could render the theory cosmologically unacceptable. The reason for this
bold statement is the following. An initially weak current on a string 
loop is amplified as the loop contracts. The current becomes sufficiently
strong to halt the loop contraction, preventing it from decaying. A stable 
state, or vorton \cite{DS87} is formed. Chiral vortons belong to the
category that is classically stable 
\cite{CM}, though their quantum mechanical stability is an open question.
The density of vortons is tightly constrained by cosmological requirements. 
For example, if vortons are sufficiently stable to survive until the present
time, then their density must be such that the universe does not become
vorton dominated. On a more conservative note, if the vortons only live
a few minutes, then their density must be such that the universe is
radiation dominated at nucleosynthesis. This enables us to constrain
theories that give rise to current-carrying strings. 

Whilst \cite{DDT1,DDT2} showed that all supersymmetric theories result
in current-carrying strings, supersymmetry breaking could destroy the zero
modes in the F-term theories. Thus, the vorton problem could solve itself
here. However, for abelian theories with symmetry breaking implemented with
a D-term, the zero mode survives, resulting in a potential vorton problem.
In a previous paper we constrained particle physics theories giving rise 
to current-carrying strings \cite{BCDT96}. The constraints applied to 
cases where the string became current-carrying at formation or else at
a subsequent phase transition, and led to fairly stringent bounds on 
current-carrying strings. However, the problem in the
chiral theory is more serious than the theories previously considered. 
This is because we previously considered a random current on the cosmic
string. However, in the chiral case, the fermion zero mode can travel in
one direction only, resulting in a maximal current and consequently a
maximum vorton density. 

In this paper we revise our previous estimates for this chiral case. Since
these cosmic strings arise in a wide class of supersymmetric and superstring
theories our results are potentially important.

\section {Cosmic Vortons}
\subsection{The case of strictly chiral vorton states.}

In a previous paper~\cite{BCDT96} we considered the generic case of 
cosmological vorton production, which is characterised by two independent
quantum numbers, $N$ and $Z$.   
The circumference of the vorton is given
approximately by
\be 
\lv\simeq(2\pi)^{1/2}\vert NZ\vert^{1/2}\mx^{-1} \ ,
\label{old 6}
\ee 
where $\mx$ is the relevant Kibble mass scale, which is defined as the
square root of the string tension $\T$ in the low current limit.
This results in the vorton energy,
\be 
\Ev\simeq\lv \mx^{\,2}\ \ . 
\label{old 5}
\ee
Since even in the general case one would expect the two quantum numbers to
have comparable magnitude, the
typical vorton mass would be 
\be
\Ev\approx N\mx\ .
\label{plus 10}
\ee
Here we will consider the more specific case where the current is strictly
chiral,
so that only one of the numbers $N$ and $Z$ can be chosen independently,
while the other will be given by the exact equality 
\be Z=N\ .
\label{chirality}
\ee

As in the earlier work~\cite{BCDT96}, the present work is carried out
within a scheme depending on two basic mass scales characterising the
temperatures of two distinct condensation processes. The first is
the temperature,
\be
\Tx\approx \mx \ ,
\label{plus 12}
\ee
at which the strings themselves are formed. The second is the comparable
or lower temperature given by
\be
\Ts\approx\ms \ ,
\label{plus 13}
\ee
where $\ms$ is the relevant carrier mass scale characterising a
current that condenses on the string in the manner described by
Witten~\cite{Witten85}. 

In the strictly chiral case~\cite{CP99}, which will be our main concern, the 
current can only be fermionic and must be  electromagnetically decoupled. 
However, whether strictly chiral or otherwise,
the requirement that the current be able to provide the
centrifugal support for an effectively stable classical vorton state
requires that the relevant vorton circumference $\lv$ should be
large compared with the Compton wavelength associated with the carrier
mass $\ms$. This requirement, namely

\be
\lv\gg \ms^{-1} \ ,
\label{minimum}
\ee
will only be satisfied if  
\be \vert NZ \vert^{1/2} \gg {\Tx\over\Ts}\ .  
\label{minprod} 
\ee

A loop that failed to satisfy this requirement would not be destined to 
survive as a vorton.

\subsection{The underlying cosmological scenario}

Like its predecessor~\cite{BCDT96}, the present analysis will be
carried out within the framework of a simple big bang theory  of the
usual type in which the universe evolves in approximate thermal
equilibrium with a cosmological background temperature $\Th$ .
The effective massless degrees of freedom at temperature $\Th$ is denoted by
$\aa$, where 
$\aa\approx 1$ at low temperatures but that in the range where vorton
production is likely to occur, from electroweak unification through to
grand unification something, like $\aa \simeq 10^2$ is a reasonable
estimate.  

All likely vorton formation processes occurred during the radiation
dominated era and ended when the universe 
dropped below the Rydberg energy scale, of the order
of $10^{-8} \GeV$ and became effectively transparent. According to the
Friedmann formula, the age $t$ of the universe is given by $t \approx H^{-1}$,
and $H$ is the Hubble parameter. During the radiation dominated era, the 
cosmological time is given by 
\be
t\approx {\mP\over\sqrt\aa\Th^2} \ ,
\label{plus 18}
\ee  
where $\mP=G^{-1/2}$ is the Planck mass.

As discussed in the following sections, any string loops present at the
epoch of the current condensation process at the temperature given by
(\ref{plus 13}) will be endowed by the ambient thermal fluctuations
with corresponding quantum numbers $N$ and $Z$ that will necessarily be
equal in the chiral case, and that subsequently, at lower temperatures,
will be conserved in between loop chopping and recombination processes
unless or until the loop length gets below the quantum stability limit
(\ref{minimum}). In the long run the loops with quantum numbers
large enough for the corresponding stability requirement (\ref{minprod})
to be satisfied will be predestined to become stationary
vortons; we refer to these as {\it protovortons}.

In the strictly chiral case the
relevant quantum numbers can be expected to be considerably larger than
in the generic case considered before. This implies that, as in the generic 
case, so a fortiori in the chiral case, whenever the carrier
condensation occurs during the friction dominated regime, a
majority of the loops already present at the temperature $\Ts$  of the
carrier condensation will qualify as protovortons. However in other 
cases (those for which the
condensation occurs at a relatively low temperature) it will only be
at a later stage that the protovorton loops get sufficiently free from
the surrounding tangle of string for subsequent quantum number changing
intersections to be negligible. The correspondingly lower
temperature, $\Tf\lta\Ts$ at which this occurs will be referred to as
the protovorton formation temperature. The protovortons will
not become vortons in the strict sense until what may be an even lower
temperature, the vorton relaxation temperature $\Tr$ say (whose value
will not be relevant for our present purpose) since the loops must
first lose their excess energy.  Whereas frictional drag and
particle production will commonly ensure fairly rapid
relaxation, there may be cases in which the only losses are due to the
much weaker mechanism of  gravitational radiation.

The raw material for vorton production is provided by the process whereby, as
the string distribution rarifies due to friction and radiation damping 
not all of its lost energy goes directly into frictional heating of the 
background or emitted radiation. Instead there will always be a certain 
fraction $\ef$ say that goes
into autonomous loops, meaning loops small enough to evolve without subsequent
collisions with the main string distribution.
Since these loops can not greatly exceed the smoothing length scale 
they will not have much fine substructure and their subsequent evolution
is very likely to satisfy the condition of avoiding quantum number changing 
self intersections as they subsequently contract. 

When the condensation occurs during the friction dominated epoch, the
autonomous loops that emerge at the condensation
temperature $\Ts$ will satisfy the condition (\ref{minprod}) and thus
be describable as protovortons. However for lower values of $\Ts$ the
majority of the loops that emerge during the period immediately
following the carrier condensation will be too small to have aquired
sufficiently large quantum numbers, and therefore will not be viable in
the long run. Nevertheless even in such unfavourable circumstances, the
monotonic increase of the damping lengthscale $\xi$ will ensure that
at some lower temperature, $\Tf<\Ts$, a later -- but therefore
less prolific -- generation of emerging loops will after all be able to
qualify as protovortons.

The small autonomously evolving loops that we refer to as protovortons
are supposed to be sufficiently small compared with the ever expanding
lengthscales characterising the rest of the string distribution 
and sufficiently smooth (due to previous
damping) to avoid destructive fragmentation by self collisions. Thus
in most individual cases, with reasonably high accuracy when
averaged, the relevant quantum numbers $N$ and $Z$ will be conserved. As a
consequence, the statistical properties of the future vorton population will
be predetermined by those of the corresponding protovorton loops at the time
of their emergence at the temperature $\Tf$.

\subsection{The smoothing lengthscale}

The scenario summarised above is based on the accepted understanding of the
Kibble mechanism\cite{V&S}, according to which, after the temperature has
dropped below $\Tx$ the effect of various damping mechanisms will remove most
of the structure below an effective smoothing length $\xi$ which will
increase monotonically as a function of time, so that nearly all the surviving
loops will be have a length $L=\oint d\ell$ that satisfies the inequality 

\be
L\gta\xi
\label{loopl}
\ee
There will thus be a distribution of string loops, of which the most numerous
will be relatively short ones, with $L\approx\xi$, that are on the verge of
emerging, or that have already emerged, as autonomous protovortons or
as loops that are about to contract and disappear, 
as the case may be. Above this smoothing length there will be a spectrum of
tangled structure, in which the number density $n$ say of closed loops and
wiggles segments extending over a radial distance greater than 
$R>\xi$ will be given by an expression of the form

\be 
n\approx {\nu\over R^3}
\label{spectr}
\ee
in which $\nu$ is a dimensionless coefficient. Note that an individual closed 
loop characterised by a radial extension
of order $R$ will typically have a much longer total random walk length
given by
\be
L\approx R^2/\xi \ .
\label{walk} 
\ee
When the string distribution is first formed, at the temperature $\Tx$, one
expects (\cite{V&S}) that the spectrum will be of the simple Brownian type
for which $\nu$ has a constant value $\nua$ say of order unity. By
causality the spectrum must always
retain a simple Brownian form for values of $R$ exceeding the horizon
lengscale $t$. The Brownian spectrum will be preserved even in the
intermediate range $\xi <R<t$ throughout the friction dominated regime.
However, as discussed below,  the situation becomes more complicated in
the subsequent regime of locally free string motion whose description
requires the use of a variable coefficient $\nu$ that will depend on both 
$R$ and $t$.

Whereas on larger scales closed loops and wiggles on very long string segments
will be tangled together, on the shortest scales characterised by the lower
cut off $\xi$, loops of a relatively smooth form lead a
comparitively autonomous existence, passing between the meshes of the ambient
tangle with only occasional collisions. It is these smallest loops that
are of interest as candidates for subsequent transformation into vortons. If
one adopts the convention that these autonomous loops are to be counted apart
from the main string distribution, the extrapolation of the spectrum
(\ref{spectr}) to wavelengths small compared with $\xi$ will be describable
by a drastic reduction of $\nu$ to a negligibly small value in this range, so
that the overall spectrum (\ref{spectr} will peak at the value $R\approx \xi$.
Alternatively, if one adopts the convention that not only wiggles but also the
autonomous small loops are to be included in the count, then there will still
be a reduction on $\nu$, but of a more moderate nature, so that instead of a
peak the overall spectrum will just have a plateau for $R\lta\xi$. 

The total number density of the small autonomous loops with length and radial
extension of the order of $\xi$ will (due to the rapid fall off of the spectrum
that is expected for larger scales) be not much less than the number density
of all closed loops (including those that exist ephemerally pending string
intersections at larger scales), and so will be given by an expression of the
form

\be
n\approx \bar\nu\  \xi^{-3}
\label{plus 22}
\ee
where $\bar\nu$ is the value of $\nu$ for values of $R$ of the order of $\xi$.
In view of the expected fall off of the spectrum, this value $\bar\nu$ will
also be interpretable as an appropriately averaged value of $\nu$. In
particular, for the Brownian case in which $\nu$ has a constant value $\nua$
over the range $R\gta \xi$, we shall evidently have $\bar\nu\approx\nua$.

\section {The Vorton Population}
\subsection{The chiral amplification effect}

The standard theory reviewed above was originally developed on the
assumption that the 
string evolution
is governed by Nambu-Goto type dynamics. However this will change
after the current carrier condensation, at the temperature $\Ts$. The
reason why the strings not only can but will carry a potentially
significant current is
 that the typical length scale $\xi$ of the string loops at the
transition temperature $\Ts$ will generally have a value, $\xis$ say,
that is considerably greater than the relevant value
$\lambda_\sigma\approx\Ts^{-1}$ of the wavelength $\lambda$
characterising the local current fluctuations induced on the string by
the thermal background.

In the generic case studied previously~\cite{BCDT96} the number
$N\approx L/\lambda$ of such fluctuation lengths around any loop of
circumference $L\gta \xi$ will overestimate the corresponding total
quantum number associated with the loop, which by a ``random walk''
argument can be expected to be only of the order of the corresponding
square root, namely $N\approx\sqrt{L/\lambda}$. However in the strictly
chiral case, in which only right (or left) moving null modes are allowed, 
the partially cancelling backwards
steps in the ``walk'' are not allowed, so the naive estimate $N\approx
L/\lambda$ will be valid. Both possibilities can be allowed for 
simultaneously by using the formula
\be
\vert Z\vert \approx N \approx \left({L\over\lambda}\right)^{1/i} \ , 
\label{N}
\ee
with $i=1$ in for the strictly chiral case, and $i=2$ for the generic case
that was studied previously.

Since the loop length $L$ is bounded below by a smoothing
length $\xi$ that will be large compared with the relevant fluctuation
wavelength
\be
\xis\gg \lambda_\sigma  \ ,
\label{plus 24}
\ee
when the current first condenses, it can be seen that even in the generic case
$i=2$, and a fortiori in the strictly chiral case $i=1$, the quantum
numbers provided by the formula (\ref{N}) will be large compared with unity.
A typical loop at that time will be characterised by
\be
L\approx\xis\ 
\label{typil}
\ee
and
\be
\lambda\approx\Ts^{-1}
\label{lambda}
\ee
so that one obtains the estimate
\be
\vert Z\vert \approx N \approx \Big(\xis\Ts\Big)^{1/ i} \ . 
\label{plus 29}
\ee

For current condensation during the friction dominated regime  this will 
always be enough to fulfil
the requirement (\ref{minprod}), but this  condition, namely
\be \xis\gg{\mx^{\, i}\over \Ts^{\,i+1}}\ ,
\label{plus 29a}\ee
will not hold for condensation later on in the radiation damping regime. 
In the latter case, typical small loops that free themselves from the
main string distribution at or soon after the time of current condensation
will ultimately disappear, since they acquire their autonomy too soon to be
viable in the long run. However there will always be a minority of longer
loops, namely those exceeding a minimum given by
\be
L\approx {\mx{\,^i}\over \Ts^{\,i+1}} \ ,
\label{longenuf}
\ee   
for which (\ref{minprod}) will be satisfied.

The condition (\ref{longenuf}) is still not quite
sufficient to qualify all of these loops as protovortons since such excepionally long loops
will be very wiggly and collision prone. In such cases it is not until a
later time at a lower temperature $\Tf$ that free protovorton loops will
emerge. By this stage, instead of its original value (\ref{lambda}), the
typical wavelength of the carrier field will be given by an expression of the
form 

\be
\lambda\approx \zf\Ts^{-1}
\label{lambdaf}
\ee
which involves a blueshift factor $\zf$ whose value is not immediately obvious
but is needed to allow for the net effect on the string of weak stretching
due to the cosmological expansion and stronger shrinking due to wiggle damping
during the intervening period as the temperature cools from $\Ts$ to $\Tf$.

In the earlier friction dominated regime $\Tf$ is identifiable with $\Ts$ so
the problem does not arise, and trivially $\zf=1$.  In the radiation damping 
era, for which the
evaluation of $\zf$ is needed, cosmological stretching  will in fact be
negligible, so obviously the net effect is that $\zf$ will be small compared
with unity. The hard part of the problem is to estimate by how much so. It is
also obvious that since, during the same period, the smoothing length $\xi$
can only increase, the corresponding final value \be L\approx \xif \label{xif}
\ee of the length of a typical loop can not be less than the
original value (\ref{typil}). It therefore follows that the new estimate
\be
\vert Z\vert \approx N \approx \left({\xif\Ts\over \zf}\right)^{1/i} \ , 
\label{Nlater}
\ee
obtained from (\ref{N}) will certainly be larger than the previous value
(\ref{plus 29}). More specifically, the value given by (\ref{Nlater}) will
increase monotonically as the value of the final temperature $\Tf$ for which
it is evaluated diminishes.

The required value of $\Tf$ -- at which the
formation of the protovorton loops occurs -- is that for which
the monotonic function (\ref{Nlater}) reaches the minimum qualifying value
given by (\ref{minprod}). This value is thus obtainable in principle by
solving the equation
\be
{\xif\over \zf} \approx {\mx^{_, i}\over \Ts^{\,i+1}}\ ,
\label{fff}
\ee
but this can only be done in practice when we have found the dependence
on $\Tf$ not only of $\xif$, which is comparatively easy, but also of $\zf$
which is rather more difficult. 

The number density of the protovorton loops when they are first formed at the
temperature $\Tf$ will be comparable with the total loop number density at the
time. By (\ref{plus 22}) it will be expressible as
\be
\nf\approx\ef\nuf\, \xif^{-3} \ ,
\label{newf}
\ee
where $\ef$ is an efficiency factor of order unity, and where $\nuf$ is the
value of the dimensionless parameter $\bar\nu$ at that time. This will also
simply have an order of unity value, $\nuf \approx \nua$ say, if the current
condenses in the friction dominated regime, for which $\Tf\approx\Ts$.
However $\nuf$ can be expected to have a lower value 
if the condensation  does not occur until later on in the
radiation dominated era. 

In all cases, since the number of protovorton loops in a comoving
volume will be approximately conserved during their subsequent damping
dominated evolution, 
then the number density $\nv$ of the resulting vortons later on at
a lower temperature $\Th$ will be given in terms of the number density $\nf$
of the proto-vorton loops at the time of condensation by
\be
{\nv\over\nf}\approx {f\over\varepsilon}\left( {\Th\over\Tf}\right)^{3} \ ,
\label{add 1}
\ee
where $f$ is a dimensionless adjustment factor that we expect to be small but
not very small compared with unity, and that will be given by
\be
f\simeq{\ef\aa\over\aaf} \ ,
\label{plus 27}
\ee
where $\aaf$ is the value of $\aa$ at the protovorton formation temperature
$\Tf$.

It follows from (\ref{plus 10}) that the corresponding mass density
will be given by 
\be
\rhv\approx N\mx \nv\ . 
\label{plus 28}
\ee 
Using the preceding estimates, the temperature dependence of the mass density
is found to be given by the general formula
\be
\rhv \approx f\nuf {\mx\Ts^{1/i}\over\zf^{1/i}\xif^{(3i-1)/i}}
\Big({T\over\Tf}\Big)^3 \ , 
\label{add 2}
\ee
in which we shall simply have $\Tf\approx\Ts$ and therefore $\zf\approx 1$
whenever (\ref{minprod}) is satisfied by (\ref{plus 29}).
It is now necessary to evaluate this in the friction and radiation damping
regimes.

\subsection{Condensation in the friction damping regime.}

If the current condensation occurs during the friction dominated epoch,
the evaluation of the quantities involved in (\ref{add 2}) is 
comparitively straightforward.
According to the standard picture\cite{KEH}, as the background temperature
$\Th$ drops below the string formation temperature $\Tx$ of the relevant
symmetry breaking phase transition, the evolution of the cosmic string network
will at first be dominated by the frictional drag of the thermal background.
It has been predicted that the relevant dynamical damping timescale $\tau$
during this period will be approximately given by 
\be
\tau\approx{\Tx^{\,2}\over\beta\,\Th^3} \ , 
\label{plus 19}
\ee
where $\beta$ is a dimensionless drag coefficient that depends on the details
of the underlying field theory but that is typically
expected\cite{{KEH},{V&V}} to be roughly of the order of unity, $\beta\approx
1$. The effect of the damping is to freeze the large scale structure, so that 
it retains the Brownian random walk form described by (\ref{spectr}) with a
fixed order of unity value $\nua$ say for the dimensionless coefficient $\nu$,
while it smooths out the microstructure  below a correlation lengthscale $\xi$
given by
 \be 
\xi\approx\sqrt{\tau t} \ ,
\label{plus 20}
\ee
where $t$ is the Hubble timescale, which is given by (\ref{plus 18}). The
required correlation lengthscale is thus finally found to be given in order of
magnitude by 
\be
\xi\approx 
\left({\mP\over\beta}\right)^{1/2}{\Tx\over\Th^{\,5/2}}\ , 
\label{old 13}
\ee
(neglecting the very weak $\aa$ dependence that would be contained in a factor
$\aa^{1/4}$,  on the assumption that this factor will not be far from unity).

The validity of the above derivation is of course based on the supposition
that the cosmological timescale $t$ is longer than $\tau$, so that the damping
process can actually be effective. It can be seen from the preceding formulae
for $t$ and $\tau$ that this condition will indeed be fulfilled just as long
as the temperature $\Th$ remains above a critical value $\Ta$ given by 
\be
\Ta \approx\frac{\Tx^{\,2}}{\beta\,\mP} \ .
\label{old 10}
\ee

So long as the condensation temperature exceeds
this critical value, i.e provided it lies in the friction drag dominated
regime
\be 
\Tx\gta \Th\gta \Ta\ , 
\label{plus 21}
\ee
it can be verified that the majority of the small string loops already present
at the time will  satisfy the condition (\ref{minprod}) for qualification as
proto-vortons, which means that the appropriate value of $L$ for substitution
in (\ref{plus 29}) will be given by (\ref{typil}). Since it can be thus seen
from (\ref{newf}) and (\ref{old 13}) that the number density of these
proto-vorton loops will be given at the time of their formation by 
\be
\nf\approx \ef\nua\left({\beta\,\Ts\over\mP}\right)^{3/2}
\left({\Ts^{\,2}\over\Tx}\right)^3\ ,
\label{plus 25}
\ee
where $\nua$ is the constant order of unity value that is retained, during the
friction dominated era, by the coefficient $\bar\nu$ in (\ref{plus 22}) and
hence by $\nuf$ in (\ref{newf}).
It follows from (\ref{add 1}) that at 
later times the number density of their mature vorton successors will be 
given in order of magnitude by
\be
\nv\approx\nua f\left({\beta\Ts\over \mP}\right)^{3/2}
\left({\Ts\Th\over\Tx}\right)^3 \ . 
\label{plus 26}
\ee

Thus, after the temperature has fallen below the value $\Tr$ 
the order of magnitude of the resulting mass density
$\rhv$ of the relic vorton population, to which the cosmic string loop
distribution will have been reduced, will be
\be
\rhv\approx \nua f N\left({\beta\Ts\over\mP}\right)^{3/2}\,
\left({\Ts\over\Tx}\right)^2 \Ts\Th^3\ . 
\label{old 17}
\ee 
Setting $\Th$ equal to $\Ts$ in (\ref{old 13}) in order to obtain the relevant
value of $\xis$, and using (\ref{plus 29}), the required
expectation value for the quantum number $N$ in the previous formula can be
estimated as
\be
N\approx\left(\Big({\mP\over\beta\Ts}\Big)^{1/2}
{\mx\over \Ts}\right)^{1/i} \ .
\label{old 21}
\ee

It follows from  (\ref{plus 10}) and (\ref{old 6}) that a typical vorton in
this relic distribution will have a mass-energy given by 
\be
\Ev\approx\left({\mP\over\beta\Ts}\right)^{1/2i}
\left({\mx\over\Ts}\right)^{1/i}\mx \ .
\label{plus 30}
\ee                    
It can thus be confirmed using (\ref{old 10}) that, as stated above, the
postulate  $\Ts > \Ta$ automatically ensures that these vortons will 
indeed satisfy the minimum length requirement (\ref{minimum}) by
a considerable margin in the strictly chiral case, though only
marginally when $\Th$ is at the lower end of this range.

For the resulting distribution of vortons, 
the mass density obtained by substituting (\ref{old 21}) 
in (\ref{old 17}), is found to be given by
\be
\rhv\approx \nua f\left({\beta\Ts\over\mP}\right)^{3/2-1/2i}
\left({\Ts\over\mx}\right)^{2-1/i}\Ts\Th^3 \ . 
\label{plus 31}
\ee

In the strictly chiral case $i=1$, this result will be expressible by
\be
{\rhv\over \Th^3}\approx {\nua f\beta\Ts^{\,3}\over\mP\,\mx} \ . 
\label{plus31}
\ee
It is to be remarked that unlike its absolute value, the relative 
augmentation factor of the mass density does not depend on the 
as yet rather uncertain efficiency factor $\p$ (which seems likely 
to be very small in the friction dominated epoch, but perhaps 
nearer the order of unity in the radiation reaction dominated epoch). 
It can be seen that in comparison with the previously studied 
generic case, $i=2$, the mass density is augmented for the 
strictly chiral case, $i=1$, by a factor that
is simply expressible as $\big(\mx/\Ts\big)^{1/2}
\big(\mP/\beta\Ts\big)^{1/4} $.
 
\subsection{Condensation in the radiation damping regime.}

For strings that may have been formed in some (non-standard, e.g.
supersymmetric) electroweak symmetry breaking transition, the scenario of the
preceding subsection is the only one that needs to be considered. However for
strings formed at much higher energies, in particular for the commonly
considered  case of GUT strings, there will be an extensive temperature range
below the Kibble transition value $\Ta$ at which friction becomes unimportant
during which the current condensation could ocurr. Since we saw that the
minimum length requirement was only marginally satisfied by typical loops when
the condensation occurred near the end of the friction dominated regime, it is
to be expected that in the cases to be considered here for which the
condensation temperature $\Ts$ occurs below the transition temperature $\Ta$
given by (\ref{old 10}) typical loops present during  the transition will not
be long enough to qualify as protovortons. This means that the vorton
formation temperature $\Tf$ will not coincide with $\Ts$ as it did in the
friction dominated regime, but that it will have a distinctly lower value,
so the scenarios to be considered in this section will be characterised by
\be
\Ta >\Ts >\Tf \ .
\label{ford}
\ee 

In such a scenario the final stage of protovorton formation will be preceeded
by a period of evolution in the temperature range $\Ta\gta \Th\gta\Tf$ during
which the effect of friction will be negligible, so that the string motion
will be effectively free, which means that the only significant dissipation
mechanism will be that of radiation reaction. Moreover during the first part
of this period, in the range $\Ta > \Th >\Ts$, the only radiation mechanism
will be gravitational, which is so weak that to begin with it will have no
perceptible effect at all, so that there will be an interval during which the
smoothing length $\xi$ remains roughly constant at the value $\xia$ it attained
at the end of the friction dominated era, which will be given, according to
(\ref{old 13}) and (\ref{old 10}) by

\be
\xia\approx {\beta^2\mP^3\over\Tx^4} \ .
\label{xicrit}
\ee
The effect of string intersections might even cause the effective smoothing
length to creep downwards slightly, but there is another weak effect with an
opposite tendency.
In an accurate analysis the global damping mechanism of the cosmological
expansion needs to be taken into account, whose effect is closely analogous
to that of friction considered in the preceding regime, though with a
damping timescale $\tau$ that is of the same order of magnitude as the Hubble
timescale $t$, so that the effect of this ``Hubble damping'' is only marginal.

During the last stage before the protovortons are formed, in the range
$\Ts>\Th>\Tf$ there will already be currents on the strings, which means that
if they are electromagnetically coupled (as is often taken for granted though
it is not necessarily the case) then the mechanism of gravitational radiation
damping may in principle be reinforced by the potentially much stronger
mechanism of electromagnetic radiation damping. However in practice even in
the coupled case, the expected currents will be too weak for this be
important, so throughout the range $\Ta>\Th>\Tf$ the gravitational radiation
is the only kind that actually matters.

The resulting gravitational smoothing scale $\xi$ will be the length of the
shortest loop for which the cosmological timescale
(\ref{plus 18}) is exceeded by the relevant radiation survival timescale,
for which the usual order of magnitude estimate has the form 
\be
t\approx {\mP^{\,2}\xi\over\Gam G\mx^{\,2}}\ ,
\label{add 5} 
\ee
where $\Gam$ is a dimensionless coefficient of order unity, and where, for
the heavy GUT strings -- the kind most commonly considered in cosmology -- the
gravitational factor will be given by $(\mx/\mP)^2\approx
10^{-6}$. The validity of the formula (\ref{add 5}) has been confirmed in many
particular cases by numerical simulations\cite{V&S}, though the value of the
coefficient turns out to be typically
\be
\Gam\approx 10^2 \ .
\label{add 6}
\ee
Equating (\ref{add 5}) to the cosmological timescale (\ref{plus 18}), the 
corresponding cut-off lengthscale can be estimated as
\be
\xi\approx{\Gam\over\sqrt\aa\,\mP}\left({\Tx\over \Th}\right)^2 \ .
\label{add 7}
\ee 

This formula (\ref{add 7}) will determine the scale of the smallest surviving
structure after the temperature has fallen below a critical value $\Tra<\Ta$
at which this value of $\xi$ becomes larger than the value (\ref{xicrit}). It
can be seen from (\ref{add 7}) that this value will be given by
\be
\Tra\approx \Big({\Gam\over\sqrt\aa}\Big)^{1/2} 
{\Tx^{\,3}\over\beta\,\mP^{\,2}} \ .
\label{Trad}
\ee
The relation (\ref{add 7}) is equivalently expressible in the form
\be
\xi\approx\kappa t
\label{defalpha}
\ee
where $\kappa$ is a constant given by
\be
\kappa\approx \Gam \Big({\Tx\over\mP}\Big)^2 \ ,
\label{alpha}
\ee
which in the case of GUT strings means $\kappa\approx 10^{-4}$.

As was discussed in more detail in the preceeding work~\cite{BCDT96},
formulae ressembling (\ref{defalpha}) have been commonly
employed in published discussions of
string simulations\cite{V&S}, but there is some confusion arising from the use
of various definitions of $\kappa$ which lead to correspondingly diverse
numerical values, usually larger than (\ref{alpha}) for the practical reason
that the simulations in question have limited resolution and in any case
neglect the gravitational reaction mechanism from which (\ref{alpha}) is
derived. To avoid ambiguity we can of course simply use the formula
(\ref{defalpha}) as a defining relation to specify the parameter $\kappa$
during the Hubble damping ``doldrum" regime $\Ta\gta\Th\gta\Tra$ (to which
most of the simulations are restricted), that is to say after friction has
become negligible  but before radiation reaction has had time to be effective,
but with this convention $\kappa$ will have to be considered as a function of
time, starting with unit value, $\kappa\approx 1$ at the begining of the
``doldrum" regime and decreasing to the very low value (\ref{alpha}) at which
it levels off after the end of this transition regime at $\Th\approx \Tra$.

In so far as the quantity $\nu$ in (\ref{spectr}) is concerned,
it can be expected that, although it will have a more complicated 
transitional behaviour  near the
lower cut off value $R/t\approx \kappa$ where $\kappa$ is the gravitational
damping constant given by (\ref{alpha}), it is resonable to expect that
in the intermediate range $\kappa\ll R/t \ll 1$ it should be given by a 
simple power scaling law of the form
\be
\nu\approx\nua \Big({R\over t}\Big)^\q \ ,
\label{scaling}
\ee
for constant values of the index $\q$ and the coeficient $\nua$, where the
latter must necessarily be identified with the order of unity constant
$\nua$ that characterised the earlier, friction dominated regime, in
order to ensure continuous matching at the horizon scale, $R/t\approx 1$,
above which the Brownian description will still prevail.  It is shown in
(\cite{TB}) that in the radiation dominated regime with which we are 
concerned here there are reasons to expect that the appropriate value of
the index should be close to but perhaps slightly greater than a lower
limit given by
\be
\q={3\over 2} \ .
\label{index}
\ee
(The analogue for the matter era in which we are situated today is a value
slightly greater than a lower limit given by $\q=2$.) 

Assuming that the formula (\ref{scaling}) still gives the right order of
magnitude at the lower end of its range (where for higher accuracy a less
simple fomula would be needed) the averaged value $\bar\nu$ in the formula
(\ref{plus 22}) for the number density $n$ of small autonomous loops will be
given in the radiation damping dominated regime $\Th\lta\Tra$ by the constant
value
\be
\bar\nu \approx \nua\, \kappa^\q
\label{barnu}
\ee
with $\kappa$ given by $(\ref{alpha})$, which means that the corresponding
value of the loop number density itself will be given according to
(\ref{defalpha}) by
\be
n\approx \nua\, \kappa^{\q-3} t^{-3} \ .
\label{nloop}
\ee

Having thus obtained a reasonably plausible provisional estimate for the
number density, the only thing that remains to be done to obtain all the
elements wanted for  working out the required result (\ref{add 2}) is to
obtain the value $\Tf$ of $\Th$ at which the protovorton formation actually
take place. To do this we have to solve the equation (\ref{fff}) that results
from the minimum length requirement, which can be seen from (\ref{add 7}) to
reduce to the simple form
\be
\zf\Tf^{\,2}\approx{\kappa\mP\over\sqrt\aa}{\Ts^{\,i+1}\over\mx^{\,i}} \ .
\label{quation}
\ee 
However before we can solve this deceptively simple equation in practice, 
we need to know the $\Th$ dependence of the factor $\z$. According to 
the reasonning described in the preceeding work~\cite{BCDT96}, this factor 
can be expected to be given by an expression of the form
\be
\z\approx\Big({\Th\over\Ts}\Big)^{1-\p} \ .
\label{blue}
\ee
where $\varepsilon$ is a dimensionless loop production efficiency factor
somewhere in the range $0 <\varepsilon\lta 1$.
It is to be remarked that if the efficiency $\p$ of loop production
were zero, this would mean that the carrier field would be blue shifted
by a factor that would be just the inverse of that by which the background
radiation is redshifted. In practice, the fairly high loop production that
is expected in the radiation damping regime
implies that although there will still be a blueshift it
will only be by a much more moderate factor.

The formula (\ref{blue}) provides what we need in principle to solve the
equation (\ref{quation}) to obtain the value $\Tf$ of $\Th$ at which the
proton loops in which we are interested actually form. In practice, assuming
(as is necessary for the scaling hypothesis to be justified) that there is no
major phase transition significantly affecting the the particle number
weighting factor $\aa$ characterising the cosmological background in the
temperature range under consideration -- so that it can be taken to have the
fixed value $\aas$ -- the solution is conveniently obtainable in the 
explicit form
\be
{\Tf\over\Ts}=\left({\kappa\mP\over\sqrt\aas\,\Ts}\Big({\Ts\over\mx}
\Big)^i\right)^{1/(3-\p)} \ .
\label{solution}
\ee

For the resulting distribution of vortons, typically having the minimum 
size compatible with the criterion (\ref{minprod}) which gives
\be\Ev\approx {\mx^2/\Ts}\ ,\label{mins}\ee 
the estimates (\ref{add 7}) and (\ref{barnu}) can be used 
in conjunction with (\ref{solution}) to evaluate the formula 
(\ref{add 2}) so that the mass density of the vorton distribution is 
finally found to be given by
\be
\rhv\approx f\nuf\left({\sqrt{\aas}\mP\over\Gam\mx}\right)^{2-\p/(3-\p)}
\left({\Ts\over\mx}\right)^{(3i-\p)/(3-\p)}\Ts\Th^3
\label{fmassden} 
\ee
or equivalently
\be
{\rhv\over\Th^3}\approx\ef\aa\nua\Gam^{-1/2}
\Big({\sqrt{\aas}\mP\over\Gam\mx}\Big)^{-\p/(3-\p)}
\Big({\Ts\over\mx}\Big)^{(3i-\p)/(3-\p)}
\Big({\mx\Ts\over\mP}\Big)\ .
\label{den 3}\ee 

For the strictly chiral case $i=1$ this expression reduces to
\be
{\rhv\over\Th^3}\approx\ef\aa\nua\Gam^{-1/2}
\Big({\sqrt{\aas}\mP\over\Gam\mx}\Big)^{-\p/(3-\p)}
\Big({\Ts^2\over\mP}\Big)\ .
\label{den3}\ee 
It can be see that in comparison with the previously studied generic
case, $i=2$, the result for this strictly chiral case is
augmented by a factor $\big(\mx/\Ts\big)^{3/(3-\p)}$.
 
It is to be remarked that whereas in the case of condensation in the
friction dominated epoch the augmentation is attributable to an
increase in the typical mass (\ref{plus 30}) of the vortons, whose number
density is not affected, on the other hand in the case of condensation 
in the radiation reaction dominated regime the augmentation is 
attributable to an increase in the number density of the vortons,
whose typical mass (\ref{mins}) is unaffected.

\section{The Nucleosynthesis Constraint.}

One of the most robust predictions of the standard cosmological model is the
abundances of the light elements that were fabricated during primordial
nucleosynthesis that occurred when the background temperature had a value,
$\TN$ say, that is given roughly by the energy needed for tunneling through
the Coulomb barrier between two protons, by

\be
\TN\approx e^4 \mp \approx 10^{-4} \GeV 
\label{plus 32}
\ee
An essential ingredient of nucleosynthesis calculations is the current
expansion rate of the universe, as determined by the cosmological background
density $\rho$, which, in the accepted picture, was at that time still
strongly radiation dominated, so that it would have had an order of magnitude
$\rhN$ given by  

\be
\rhN \approx \aa\TN^{\,4} \ .  
\label{plus 33}
\ee

In order to preserve this well established picture, it is necessary
that the vorton distribution should satisfy $\rhv\lta\rhN$
when $\Th\approx \TN$, which for the case of carrier condensation
in the friction dominated epoch is expressible as the condition
\be
 {\ef\nua\beta\Ts^3\over\aas\mP\, \mx\TN} 
\lta 1 \ . 
\label{old 24}
\ee

The condition of condensation in the friction dominated regime will
automatically be satisfied in the important case for which the carrier
condensation occurs immediately after the string forming phase
transition itself, i.e. for which one has $\Ts\approx \mx$.
In this case (\ref{old 24}) gives 
\be \Ts\lta\left({\aas \mP\,\TN\over\ef\nua\beta}\right)^{1/2}
 \label{old 25}
\ee
Substituting the expression (\ref{plus 32}) for $\TN$ into this, taking $\aas
\approx 10^2$ and assuming that the
factor $(\ef\nua)$ and the drag factor $\beta$ are
of the order of unity yields the inequality 
\be
\Ts\lta\ e^2\big(\aas\mP\, \mp\big)^{1/2}
\approx 10^8\ \GeV \  
\label{plus 34}
\ee
as the condition that must be satisfied by the formation temperature of 
{\it cosmic strings in which a strictly chiral current condenses
immediately}, subject to the rather
conservative assumption that the resulting vortons last for at least a few
minutes. It is to be observed that this condition is only marginally
more severe than what was obtained for the non chiral generic 
case~\cite{BCDT96} and that while it rules out the formation of
such strings during any conceivable GUT transition, it is consistent by a
wide margin with their formation at temperatures close to that of the 
electroweak symmetry breaking transition.

If instead of supposing that the current condensation was immediate,
we now suppose that the strings were formed at the GUT transition, i.e.
$\mx\approx\TGUT\approx 10^{16}$ GeV, but that the condensation temperature $\Ts$ is very
much lower than this, then the analogous limit is obtained for
a value of $\Ts$ that lies in the radiation reaction dominated range
governed by (\ref{den3}) so that as the analogue of (\ref{old 24})
we shall obtain
\be
\ef\nua\Gam^{-1/2}
\Big({\sqrt{\aas}\mP\over\Gam\TGUT}\Big)^{-\p/(3-\p)}
\Big({\Ts^2\over\mP\,\TN}\Big)\lta 1\ .
\label{pplus 26}\ee 

To find the highest temperature at which GUT strings can become
chirally conducting without violating the nucleosynthesis constraints
we now set
$\Tx$ equal to $\TGUT$ in (\ref{old 24}). Since $\p$ is small compared
to 3 its effect in the index can be neglected. We thereby obtain
\be\Ts\lta \left( \Big({\Gam^{1/2}\mP\, \TN\over\ef\nua}\Big) 
\Big({\Gam\TGUT\over \sqrt{\aas}\mP}\Big)^{-\p/(3-\p)}\right)^{1/2}
\approx 10^9 \GeV\label{old 26}\ee
where, in the last step, we have used the estimates
$\aas\approx \Gam\approx  10^2$ and $\TGUT\approx 10^{-3}\mP$ Gev,
and have neglected the dependence on the rather uncertain but 
supposedly order of unity quantities $(\ef\nua)$ and $\beta$. 
This limit is considerably lower (by a factor of the order of a thousand)
than the analogous limit~\cite{BCDT96} for the generic non chiral case,
and consistently with what was postulated in its derivation,
it can be seen to be well within the radiation reaction dominated range.

The upshot is that subject again to the rather conservative assumption 
that the resulting vortons survive until the time of element formation,
which occurs within only a few minute, a theory in which GUT cosmic 
strings  become chirally conducting above $10^{9}$ GeV is inconsistent 
with the observational data.

\section{The Dark Matter Constraint.}

We now consider the rather stronger constraints that can be obtained if at
least a substantial fraction of the vortons are sufficiently stable to last
until the present epoch. On the basis of the standard (Einstein) theory of
gravity, it is generally accepted that the virial equilibrium of galaxies and
particularly of clusters of galaxy requires the existence of a cosmological
distribution of ``dark" matter with density considerably in excess of the
baryonic matter that is directly observed, mainly in the form of stars, with
density $\rhb\approx 10^{-31}$ gm/cm$^3$. On the other hand, on the same
basis, it is also generally accepted that to be consistent with the
formation of structures such as galaxies, starting from initially small
inhomeneneities in an approximately homogeneous background, it is necessary
that the total amount of this ``dark" matter should not greatly exceed the
critical closure density, namely
\be
\rhc\approx 10^{-29} {\rm gm \ cm^{-3}}
\label{add 15}
\ee
(which is about $2\times 10^{-123}$ in dimensionless Planck units).
Extrapolating back, as a function of the cosmological temperature $\Th$, this
maximum cold dark matter contribution will scale like the entropy density so
that it will be given by the expression
\be
\rhc\approx \aa\mc\Th^3\ ,
\label{plus 35}
\ee
where $\mc$ is a constant mass factor. Since $\aa\approx1$ at the present
epoch, the required value of $\mc$ (which is roughly interpretable as the
critical mass per black body photon) can be estimated as
\be
\mc\approx 10^{-28}\mP\approx 1\ \hbox{eV}\  .
\label{plus 36}
\ee

The constraint to which this dark matter limit leads is expressible as
\be
\Ov \equiv {\rhv\over\rhc}\lta 1\ .
\label{plus 38}
\ee

In the case of vortons formed as a result of condensation during the 
friction damping regime, which by (\ref{old 10}) requires
\be
\beta\,\mP\,\Ts\gta \Tx^{\,2} \ ,
\label{add 16}
\ee
the relevant estimate for the vortonic dark matter fraction is obtainable
from (\ref{plus31}) as
\be
\Ov\approx {\nua f\beta\Ts^3\over\mP\, \mx\, \mc} \ . 
\label{old 27}
\ee

In particular this formula applies to the case $\Ts\approx \mx$
in which the carrier
condensation occurs very soon after the strings themselves are formed, 
for which one obtains
\be \Ts\lta \Big({\mP\, \mc\over \nua f}\Big)^{1/2}\ . \approx 10^5\GeV
\label{plus 40}\ee

However if we are concerned with 
the general category of strings formed at the GUT level, then we shall be
obliged to consider the case of vortons formed as a result of condensation
during the gravitational radiation damping regimes
\be
\Ts\ll{\Tx^2\over\beta\mP}
\label{add 17}
\ee
for which by (\ref{den3}), the relevant estimate for the vortonic
dark matter fraction is obtained as                                 
\be
\Ov\approx\ef\aa\nua\Gam^{-1/2}
\Big({\sqrt{\aas}\mP\over\Gam\mx}\Big)^{-\p/(3-\p)}
\Big({\Ts^2\over\mP\, \mc}\Big)\ .
\label{add 18}\ee 
 Setting $\Tx$ equal to $\TGUT$ in
this formula, and dropping the order of unity coefficients in view of the
low indices involved, we obtain the corresponding limit as
\be 
\Ts\lta\left(\mP\,\mc \Big({\Gam\TGUT\over\sqrt{\aas}\mP}
\Big)^{-\p/(3-\p)}\right)^{1/2}\approx 10^5 \GeV
\ .
\label{add 19}
\ee
where, as for the analogous nucleosynthesis limit (\ref{old 26}),
the last step is based on use of the estimates
$\aas\approx \Gam\approx  10^2$ and $\TGUT\approx 10^{-3}\mP$ Gev,
and on neglect of the dependence on the rather uncertain but 
supposedly order of unity quantities $(\ef$ and $\nua$). 

\section{Conclusions}

In this paper we have constrained particle physics theories that give
rise to chiral current-carrying strings. Such chiral strings arise
in a class of supersymmetric theories where the symmetry is broken
with a D-term. Our bounds apply to all such theories, including those
derived from superstring theories. We have derived these bounds from
from cosmological considerations, requiring that the resulting stable
loops, or vortons, do not dominate the energy density of the universe.
We have considered two possibilities. Either the vortons live only a
few minutes, in which case, we demanded that the universe be
radiation dominated at the time of nucleosynthesis. If the vortons
are stable enough to survive to the present time, then we demanded
that they donot overclose the universe. 

The specially simple dynamical behaviour~\cite{CP99} of
strictly chiral strings and the related consideration that their
vorton states will
have only half as many degrees of freedom of internal excitation,
and therefore only half as many conceivable decay modes as in
the case of generic non-chiral Witten currents means that
these strictly chiral vortons can be expected to be very much more
stable than most other kinds. It therefore seems highly plausible
that they would survive not merely to the time of nucleosynthesis
but to the present cosmological epoch, which means that it is
the rather severe limits (\ref{plus 40}) and (\ref{add 19}) that
are relevant. In both of these cases the upper limit for the energy
of the chiral current forming phase transition (whether the strings
themselves were formed in the same transition or much earlier
at the GUT transition) is found to be of the order of $10^5$ GeV,
which happens to be near the upper limit of the range considered
likely to characterise the electroweak phase transition.

The preceeding conclusion has both negative and positive
implications. On the negative side it seems to exclude theories
in which chiral string currents are formed at energies much larger
than the electroweak level. On the positive side it suggests
that vortons supported by strictly chiral currents that condensed
during, or just above, the electroweak phase transition might conceivably form
a significant fraction of the dark matter in the universe. Such
considerations could well be relevant to superstring theories with large
extra dimensions.

\section {Acknowledgements}
This work was supported in part by an ESF network. We thank the organisers
of the Les Houches and Dresden meetings for providing a stimulating 
atmosphere. The work of ACD was supported in part by PPARC.

\end{document}